\documentclass{jpp}
\usepackage{graphicx}
\usepackage{hyperref}

\usepackage[utf8]{inputenc}
\usepackage[T1]{fontenc}
\usepackage{amsmath}

\usepackage{orcidlink}

\shorttitle{Variational Quantum Simulation of the Fokker-Planck Equation}

\shortauthor{Ó. Amaro, L. I. Iñigo Gamiz, and M. Vranic}

\usepackage{comment}
\usepackage{appendix}

\title{Variational Quantum Simulation of the Fokker-Planck Equation applied to Quantum Radiation Reaction}

\author{
    Óscar Amaro \orcidlink{0000-0003-0615-0686} \aff{1} \corresp{\email{oscar.amaro@tecnico.ulisboa.pt}},
    L. I. Iñigo Gamiz \orcidlink{0000-0002-7393-0896} \aff{1},
    M. Vranic \orcidlink{0000-0003-3764-0645} \aff{1}
  }

\affiliation{\aff{1}GoLP/Instituto de Plasma e Fus\~ao Nuclear, Instituto Superior T\'ecnico, Universidade de Lisboa, Lisbon, Portugal}

\begin{document}

\maketitle

\begin{abstract}
    Near-future experiments with Petawatt class lasers are expected to produce a high flux of gamma-ray photons and electron-positron pairs through Strong Field Quantum Electrodynamical processes.

    Simulations of the expected regime of laser-matter interaction are computationally intensive due to the disparity of the spatial and temporal scales and because quantum and classical descriptions need to be accounted for simultaneously (classical for collective effects and quantum for nearly-instantaneous events of hard photon emission and pair creation). A typical configuration for experiments is a scattering of an electron and a laser beam which can be mapped to an equivalent problem with a constant magnetic field.

    We study the stochastic cooling of an electron beam in a strong, constant, uniform magnetic field, both its particle distribution functions and their energy momenta.
    We start by obtaining approximate closed-form analytical solutions to the relevant observables.
    Then, we apply the quantum-hybrid Variational Quantum Imaginary Time Evolution to the Fokker-Planck equation describing this process and compare it against theory and results from Particle-In-Cell simulations and classical Partial Differential Equation solvers, showing good agreement.
    This work will be useful as a first step towards quantum simulation of plasma physics scenarios where diffusion processes are important, particularly in strong electromagnetic fields.
\end{abstract}

\section{Introduction}

Strong-field quantum electrodynamics (SFQED) studies the interaction between matter and intense electromagnetic fields. In recent years, there has been a growing interest in this area due to the availability of high-intensity laser sources, which enable the exploration of novel physical phenomena, with experiments being planned for the near future at: 
 \cite{ELI}, Apollon \cite{HPLSEPapadopoulos2016}, CoReLS \cite{OEYoon2019}, \cite{FACET-II,meurenSeminalHEDPResearch2020}, LUXE \cite{abramowiczLetterIntentLUXE2019, abramowiczConceptualDesignReport2021}, \cite{EXCELS}, \cite{ZEUS}, \cite{OMEGA}, \cite{HIBEF}, among others. 

Many of the proposed experimental setups feature scattering of intense, focused laser pulses with either relativistic electron beams or high-energy photons.

Some experimental evidence for radiation reaction has been observed \cite{coleExperimentalEvidenceRadiation2018, poderExperimentalSignaturesQuantum2018, losObservationQuantumEffects2024}.
However, for more detailed tests of the models applied in this field, additional theoretical and numerical studies are required to understand the effect of ``non-ideal'' experimental conditions, such as laser jitter, probe beam emittance, synchronization of the collision, etc. In the future, better control over these parameters will allow precision studies of radiation reaction, that is, the recoil on the charged particles that emit high-energy photons and electron-positron production in the lab, among other processes. These can then be applied in physics models of plasmas in extreme astrophysical environments, such as black holes and pulsar magnetospheres \cite{uzdenskyPlasmaPhysicsExtreme2014, timokhinTimedependentPairCascades2010, medinPairCascadesMagnetospheres2010, cruzCoherentEmissionQED2021, schoefflerHighenergySynchrotronFlares2023a}.

While current laser technology allows us to test strong-field plasma physics in a ``semi-classical'' regime, as peak intensity continues to increase, we need to start to explore the fully nonperturbative, quantum dynamics of fermions, high-energy photons and the laser field. This regime is expected to require first-principles simulation techniques, thus putting constraints on the more standard particle-in-cell simulations, even if including Monte-Carlo routines for the quantum processes.

Quantum Computing has the potential to handle the complexity of the many-body-physics dynamics in these extreme plasmas. Recently, the Plasma Physics community has started to adapt standard plasma setups and theory to the quantum algorithmic framework \cite{dodinApplicationsQuantumComputing2021, josephQuantumComputingFusion2023, amaroLivingReviewQuantum2023}. However, the intersection between Plasma Physics, Quantum Field Theory, and Quantum Computing remains unexplored.

In this work, we aim to address some of these questions, first by deriving approximate particle distributions and benchmarking with the particle-in-cell code \texttt{OSIRIS} \cite{fonsecaOSIRISThreeDimensionalFully2002}, that describes QED effects with Monte-Carlo and has already been tested in this regime \cite{vranicQuantumRadiationReaction2016}, and later by applying a hybrid quantum algorithm to the simulation of a Fokker-Planck equation relevant for SFQED. We choose the simplest possible field configuration, a strong constant magnetic field background, that could be generalised in the future.
We start with a relativistic electron beam, which can propagate and lose energy to radiation. We follow the evolution of the electron distribution function over time, accounting for the stochastic nature of the quantum emission.

This manuscript is structured as follows.
In section \ref{sc:analytical}, we describe the setup of an electron beam propagating perpendicularly to a strong magnetic field and radiating energy in the form of photons in the so-called semi-classical regime. We introduce the Fokker-Planck equation and derive solutions for the evolution of the distribution function and the first two energy moments, with details in appendix \ref{appendix:fokkerplanck}.
In section \ref{sc:variational}, we describe the quantum variational simulation of the Fokker-Planck equation, the choice of ansatz, and the numerical evolution of the parameters using Variational Quantum Imaginary Time Evolution (VarQITE).
Conclusions are given in section \ref{sc:conclusions}.

\section{The Fokker-Planck equation for Quantum Radiation Reaction}\label{sc:analytical}

In this section, we introduce the main parameters for the physical setup, the Fokker-Planck equation describing the interaction of electrons with an intense magnetic field, and approximate formulas for the first two moments of the electron distribution functions.
The importance of quantum effects of relativistic particles in strong fields is controlled by the quantum nonlinearity parameters, which for leptons and photons are given by

\begin{equation}
    \chi \equiv \dfrac{\sqrt{ (p_\mu ~F^{\mu \nu})^2 }}{E_S ~m c}, \quad \chi_\gamma \equiv \dfrac{\sqrt{ (\hbar k_\mu ~F^{\mu \nu})^2 }}{E_S ~m c}
\end{equation}

\noindent where $m$ is the electron mass, $c$ is the speed of light in vacuum, $E_S = m^2 c^3/(e\hbar)= 1.32 \times 10^{18}$ V/m and $B_S = E_S/c= 4.41 \times 10^{9}$ T represent the critical Schwinger electric and magnetic fields, $e$ is the elementary electric charge, $\hbar$ is the reduced Planck constant, $F_{\mu \nu}$ is the electromagnetic field tensor, and $p_\mu, k_\mu$ are the leptonic and photonic 4-momenta.
In the case of a relativistic electron moving perpendicularly to a strong, uniform and constant magnetic field, we have

\begin{equation}
    \chi = \gamma B/B_S
\end{equation}


The electron distribution function $f=\mathrm{d}N/\mathrm{d}\gamma$, that is, the number of electrons $\mathrm{d}N$ per interval of energy $\mathrm{d}\gamma$, evolves through 
a Fokker-Planck equation describing stochastic energy losses

\begin{equation}
    \frac{\partial f(t, \gamma)}{\partial t}=\frac{\partial}{\partial \gamma}\left [-(A f)+\frac{1}{2} \frac{\partial}{\partial \gamma}\left(B f\right)\right]
    \label{eq:fokker}
\end{equation}

This is the Partial Differential Equation that will be simulated through a quantum algorithm in a later section. The scalar drift and diffusion space-dependent terms can be approximated in the $\chi \ll 1$ regime as

\begin{equation}
    \begin{split}
        A \sim \dfrac{2}{3} \dfrac{\alpha m c^2}{\hbar} \chi^2 = a~ \gamma^2, ~\mathcal{B} \sim \dfrac{55}{24 \sqrt 3} \dfrac{\alpha m c^2}{\hbar} \gamma ~ \chi^3 = \dfrac{b^2}{2} ~ \gamma^4
    \end{split}
    \label{eq:FP_AB_app}
\end{equation}

\noindent where $a \equiv 2\alpha k^2/(3 \tau_c)$, $\alpha\equiv e^2/(\hbar c)$ is the fine structure constant, and $b \equiv \sqrt{55\alpha k^3 /(12\sqrt{3}\tau_c)}$, with Compton time $\tau_c \equiv \hbar/(mc^2)$, and normalized magnetic field $k \equiv B/B_S$ \cite{neitzStochasticityEffectsQuantum2013, vranicQuantumRadiationReaction2016}.
In general, there are no closed-form solutions for the electron distribution function valid in all regimes of quantum nonlinearity $\chi$. 
Further details on the theory of Quantum Radiation Reaction are presented in appendix \ref{appendix:fokkerplanck}.

\subsection{Evolution of the electron distribution function}

In this section, we use the classical approach of simulating the particle motion coupled with Monte-Carlo to account for the hard-photon emission and quantum radiation reaction. As such, it can describe the evolution of the electron distribution, including the energy loss and quantum stochasticity-induced widening of the spectrum.

In the $\chi\ll1$ regime, an initially narrow distribution with energy $\gamma_0$ retains an approximately Gaussian shape throughout the interaction. Following the approach of  \cite{torgrimssonQuantumRadiationReaction2024,torgrimssonQuantumRadiationReaction2024a,blackburnAnalyticalSolutionsQuantum2024}, which is based on a perturbative expansion in $\chi$,  the mean energy $\mu= \langle \gamma \rangle = \int \gamma f~\mathrm{d}\gamma$ and energy spread $\sigma^2 = \int (\gamma-\mu)^2 f~\mathrm{d}\gamma$ in a constant magnetic field can be obtained as

\begin{equation}
    \dfrac{\langle \gamma \rangle}{\gamma_0} \sim \dfrac{1}{1+2 R_c t/3} +  \dfrac{165\chi_0}{8 \sqrt{3}(1+2/3R_ct)^2}  \log\left(1+\dfrac{2R_ct}{3}\right)
    \label{eq:Blackburn_average}
\end{equation}

\begin{equation}
    \dfrac{\sigma^2}{\gamma_0^2} \sim  \dfrac{\sigma_0^2 + 55 R_c \chi_0 t/(24 \sqrt{3})}{(1+2R_c t/3)^4}
    \label{eq:Blackburn_spread}
\end{equation}

\noindent where $R_c \equiv \alpha ~b_0 ~\chi_0$ the classical radiation reaction parameter, and $b_0 \equiv e B/(m~\omega_c)$ is an adimensional normalized magnetic field, and $\omega_c \equiv e B/(m \gamma_0)$ is the synchrotron frequency. In our simulations, the time is normalized as $t\rightarrow t~ \omega_c$, and the average electron energy and normalized magnetic field are $\gamma_0=b_0=1800$. From equation \ref{eq:Blackburn_spread}, the maximum energy spread occurs approximately at $t\sim 1/(2 R_c)$ for $\sigma_{max} \sim 3^{5/4} \sqrt{55}/64 ~\gamma_0 \sqrt{\chi_0}$, which recovers the scaling derived in \cite{vranicQuantumRadiationReaction2016}. 

To test the validity of these analytical expressions, we run 1D3V simulations with an electron beam moving perpendicular to the constant, uniform magnetic field vector. The timestep is $\mathrm{d}t = 0.001 ~\omega_c^{-1}$, the cell size $\mathrm{d}x=0.049~c/\omega_c$, and the normalizing frequencies and external magnetic field values are chosen to enforce $\chi_0=\{10^{-3},10^{-2}, 10^{-1}\}$.

\begin{figure}
    \centering
    \includegraphics[width=300pt]{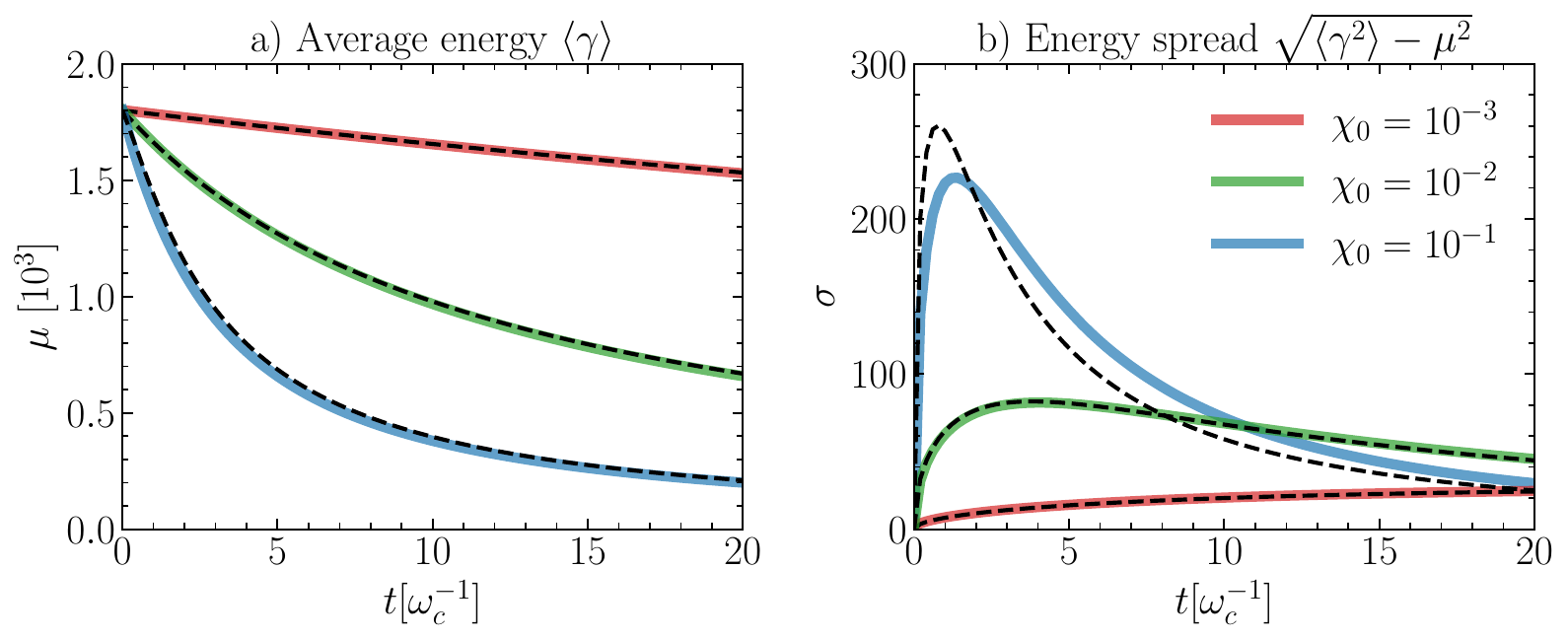}
    \caption{a) Average energy and b) spread: \texttt{OSIRIS} simulations (full, color line) vs theory (dashed lines). For the $\chi_0=10^{-1}$ results, the time axis was multiplied by $2$ to be visible in the same range as the other values.}
    \label{fig:analytical_moments}
\end{figure}

In figure \ref{fig:analytical_moments}, we show results from \texttt{OSIRIS} Monte-Carlo simulations for different values of the external magnetic field (different values of initial, average $\chi_0$), and the moments from equations (\ref{eq:Blackburn_average}) and (\ref{eq:Blackburn_spread}). The latter expressions capture the simulation results well, with some loss of accuracy for $\chi_0=10^{-1}$.
For the curve $\chi_0=10^{-1}$, the time axis is scaled ($\times2$), for better readability.

\begin{figure}
    \centering
    \includegraphics[width=300pt]{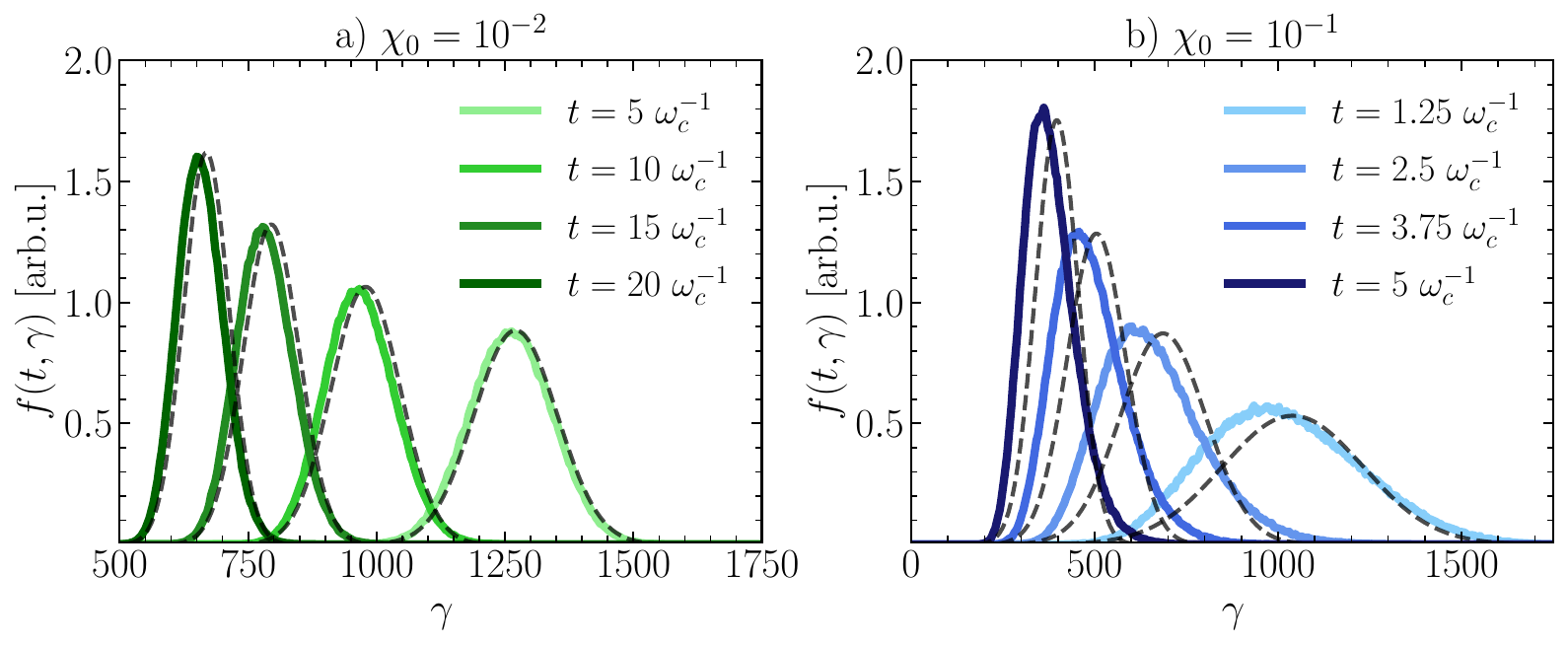}
    \caption{Snapshots of the distribution functions, taking the analytical formulas for the mean energy and energy spread (equations (\ref{eq:Blackburn_average}) and (\ref{eq:Blackburn_spread})) for $\chi_0=\{10^{-2}, 10^{-1}\}$. Dashed lines - Gaussian approximation, colored lines - \texttt{OSIRIS} simulation results. For $\chi_0=10^{-1}$ there is a visible deviation from the Gaussian approximation.}
    \label{fig:analytical_distributions}
\end{figure}

In figure \ref{fig:analytical_distributions}, we show snapshots of Gaussian distribution functions using the parameters $(\mu,\sigma)$ from equation \ref{eq:Blackburn_average}, and energy histograms from the \texttt{OSIRIS} Monte-Carlo simulations. Although for $\chi_0=10^{-2}$ there is good agreement with simulation results, it is clear that in the higher $\chi_0=10^{-1}$ regime, the validity of the Gaussian functional approximation is lost, even though the moments $(\mu, \sigma)$ predicted analytically are close to what was obtained in the simulation. This deviation is due to an increased skewness of the distributions.
These results will serve as a ``ground truth'' for testing the quantum algorithm presented in the following section.

\section{Variational Quantum Simulation}\label{sc:variational}

In this section, we introduce the Variational Quantum Simulation (VQS) method employed in our study.

Quantum Computing holds the potential to revolutionize simulations of complex systems, encompassing both quantum many-body problems and certain classical physics phenomena. The fundamental unit of quantum computation is the qubit, which, unlike a classical bit, can exist in a superposition of states described by $|\psi\rangle = a_0|0\rangle + a_1|1\rangle$, where $a_0$ and $a_1$ are complex amplitudes satisfying the normalization condition $|a_0|^2 + |a_1|^2 = 1$ \cite{nielsenQuantumComputationQuantum2010}. When multiple qubits are combined into a register, their joint state can exhibit entanglement — a uniquely quantum mechanical correlation that leads to an exponentially large state space, challenging to simulate efficiently on classical computers.

Currently, quantum computers are in the Noisy Intermediate-Scale Quantum (NISQ) era \cite{preskillQuantumComputingNISQ2018}, characterized by devices that contain a moderate number of qubits but are susceptible to errors and decoherence. These limitations restrict the depth and complexity of quantum circuits that can be reliably executed, posing significant challenges for implementing algorithms requiring long coherence times and high gate fidelities.

To address these challenges, Variational Quantum Circuits (VQCs) have emerged as a promising approach suitable for NISQ devices. VQCs are hybrid quantum-classical algorithms that employ parameterized quantum circuits optimized using classical optimization routines to minimize a cost function, typically the expectation value of an observable. This method reduces the required circuit depth by offloading part of the computational workload to classical processors, making it more practical for current quantum hardware.
Variational quantum algorithms have found applications across various fields, including Quantum Chemistry, Material Science, Biology and others \cite{peruzzoVariationalEigenvalueSolver2014}.
By carefully designing the variational circuits and selecting appropriate ansätze, VQCs leverage the expressive power of quantum systems while operating within the practical constraints of NISQ devices. Extracting meaningful information from the exponentially large quantum state space necessitates meticulous selection and measurement of observables, as quantum measurements yield probabilistic outcomes that collapse the superposed state. The variational approach thus provides a flexible and adaptive framework for quantum simulation, contributing significantly to the advancement of quantum computing applications in science and engineering.

In Plasma Physics, several phenomena of interest are intrinsically dissipative. Since dissipation is an irreversible process, this is challenging to model on quantum computers.
In \cite{engelQuantumAlgorithmVlasov2019}, the authors studied Landau damping on a quantum framework using a linearized version of the Vlasov equation. Despite the name ``damping'', the energy of the system is not lost but rather transferred from the electric field to the particle distribution function in a reversible/unitary manner.
In \cite{vissersImplementingQuantumStochastic2019} a ``quantum stochastic'' process of a laser driven two-level atom interacting with an Electromagnetic field is studied. However, a procedure to generalize beyond two-level systems is not provided.
In \cite{kuboVariationalQuantumSimulations2021}, a Partial/Stochastic Differential Equation (PDE/SDE) solver of the Fokker-Planck equation in the form of an Itô-process is proposed in a Quantum Variational framework. This approach was then generalized in \cite{alghassiVariationalQuantumAlgorithm2022} through the Feynman-Kac formula, which unifies the heat, Schrödinger, Black-Scholes, Hamilton-Jacobi, and Fokker-Planck equations under this formalism.

The expressibility of variational quantum circuits is one of the most important figures of merit in the field of VQS. It is defined as the ability of the variational quantum circuit to produce a variety of quantum wavefunctions — the higher this value, the larger the fraction of the Hilbert space accessible through the circuit. This metric has been studied in detail, and heuristics on which architectures one can reach higher expressibility have been discussed  \cite{simExpressibilityEntanglingCapability2019}. However, few proofs or universal rules have been derived so far.
In the case of \cite{kuboVariationalQuantumSimulations2021}, the authors suggest using an ansatz with only $R_Y$ and CNOT quantum gates, which permits only real-valued amplitudes of the wavefunction, but it also allows for negative values.
In \cite{dasguptaLoadingProbabilityDistributions2022}, the authors suggest an architecture to enforce even-symmetry of the real-valued wavefunction around the middle of the computational basis representation.
%
%
In \cite{endoVariationalQuantumSimulation2020}, the motivation for using compact variational circuits is based on the intuition that the dynamics of the physical system only span low energy states and therefore is limited to a small portion of the entire Hilbert space.

In this work, we consider wave/distribution functions that are well localized in space but can have a mean position/energy which can be off-centred, that is, not located at the middle of the computational basis $|01...1\rangle, |10...0\rangle$.
For this, we use a quantum circuit which simulates the advection equation based on the finite differences operator as presented in \cite{satoHamiltonianSimulationHyperbolic2024}. This operation can be used to shift the entire wavefunction, which allows the exploration of a larger space of states.

\begin{figure}
    \centering
    \includegraphics[width=350pt]{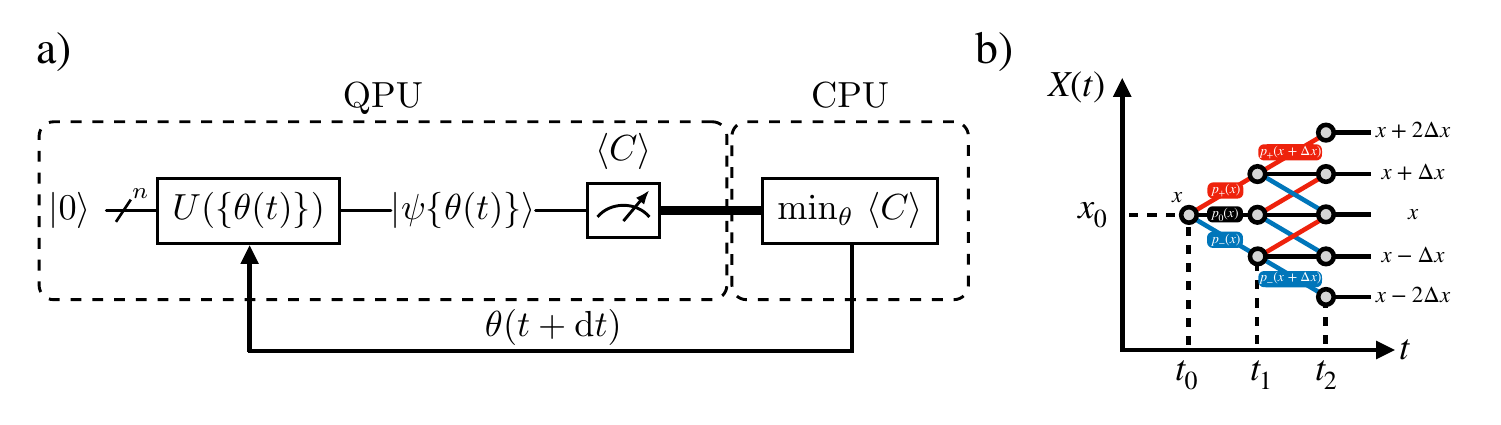}
    \caption{a) Variational quantum algorithm. From the Quantum Processing Unit (QPU) we measure a cost function, from which the variational parameters are optimized in the CPU. b) Finite Difference ``trinomial model'' exemplifying the locality of the Fokker-Planck operator (adapted from \cite{kuboVariationalQuantumSimulations2021}.}
    \label{fig:variational_alg_trinomial}
\end{figure}

In figure \ref{fig:variational_alg_trinomial}, we show the hybrid quantum optimization loop. First, at a time-step $t$, a quantum wavefunction
is produced by the quantum circuit, where some of the quantum unitary operations are parametrized by values $\theta(t)$. Then, a cost function / observable $\langle C \rangle$ is measured from the quantum circuit and minimized/updated through an algorithm run on a classical device. The new parameters are then fed-back to the quantum circuit to produce the wavefunction for the next time step $\theta(t+\mathrm{d}t)$.
The Finite-Difference approximation to the Fokker-Planck operator is of the trinomial type, where any grid cell at any time step is only influenced by its two close neighboring cells. Higher-order Finite-Difference schemes are possible to apply but require more complex algorithms.

\subsection{Variational ansatz}

\begin{figure}
    \centering
    \includegraphics[width=350pt]{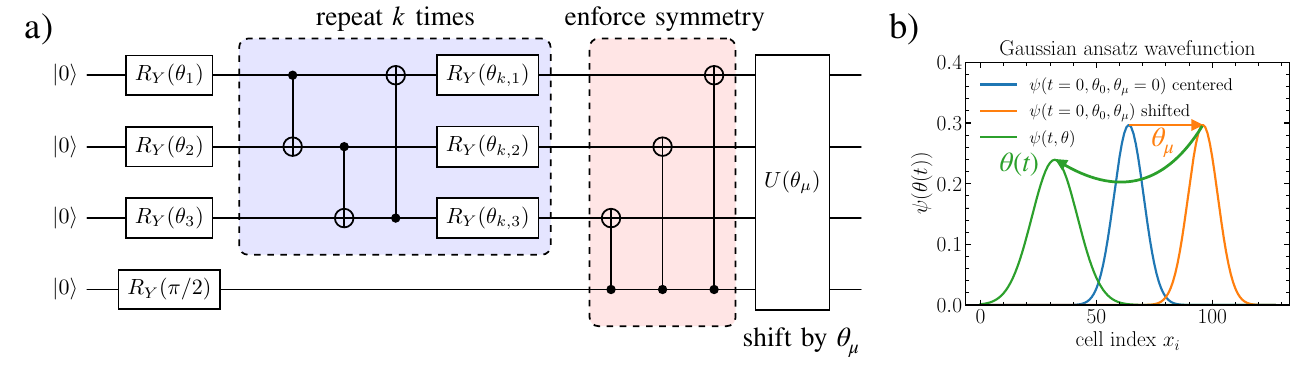}
    \caption{a) Variational ansatz used in this work. Only 4 qubits are shown for simplicity. The blue box represents a variational block that can be repeated. The red box represents the enforcing of even-symmetry on the wavefunction. In the last step, the advection operator is applied. b) Sketch of the typical evolution of the wavefunction.}
    \label{fig:ansatzdriftdiffusion}
\end{figure}

In figure \ref{fig:ansatzdriftdiffusion}, we show the ansatz used in this work. 
In the first variational block (blue), we produce a real-valued wavefunction using parameters $\bf{\theta}$ and gates $\mathrm{CNOT}-R_Y$ in a ring structure (the qubit $n-1$ connects with the first qubit). This block can be repeated to increase the expressibility of the ansatz.
The second block (red) consists of enforcing even-symmetry on the wavefunction; that is, the amplitudes have a mirror symmetry around half of the computational basis $\psi_{0111} = \psi_{1000}, \psi_{0000}=\psi_{1111}$. If, instead, we apply an angle $-\pi/2$ on the $R_Y$ gate acting on the last qubit, then the wavefunction becomes odd instead of even-symmetric. At $t=0$, we fit these parameters so that the wavefunction has an approximate Gaussian shape with a prescribed standard deviation $\sigma$. In this way, the parameters $\bf{\theta}$ are implicit functions of $\sigma$.

In the final step of the quantum circuit, we apply the advection operator $e^{-i \theta_\mu \partial_x }$, which is a unitary operation, and which translates the entire wavefunction by some quantity $\theta_\mu$. In \cite{satoHamiltonianSimulationHyperbolic2024}, the authors provide a quantum circuit for the Finite-Difference version of the 1st order of $\partial_x \sim (s_+-s_-)/2$, where $s_+ = s_-^\dagger$ is the operator that shifts the wavefunction one cell to the right (left) with a periodic boundary condition. Here, we use the 2nd order of this Finite-Difference operator to enforce higher fidelity of the translation.

\subsection{Variational Quantum Simulation of the Fokker-Planck equation}

\begin{figure}
    \centering
    \includegraphics[width=340pt]{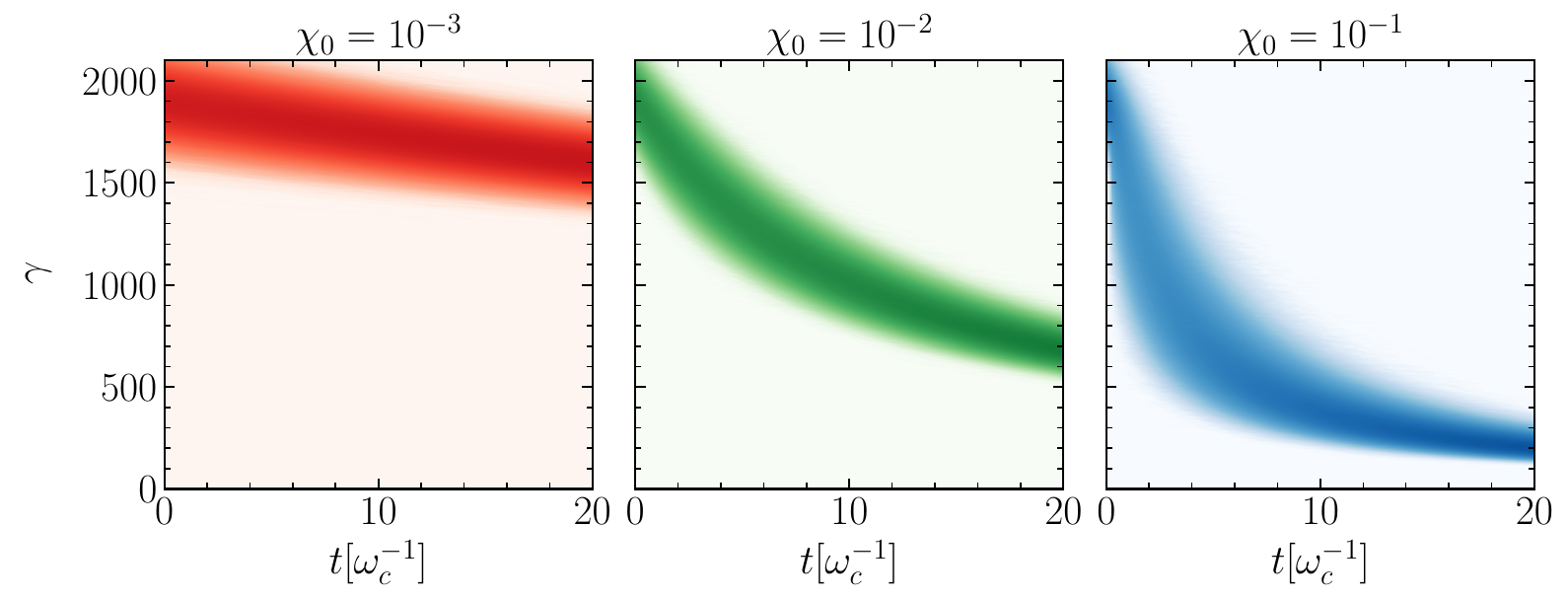}
    \caption{Evolution of electron distribution functions from \texttt{OSIRIS} simulations for different initial average $\chi_0$ values and $\sigma_0=90$.}
    \label{fig:FokkerPlanckSolution_triptych}
\end{figure}

In figure \ref{fig:FokkerPlanckSolution_triptych}, we show the complete evolution of the distribution functions for $\mu_0=1800$, $\sigma_0=90$ from \texttt{OSIRIS} Monte-Carlo simulations, following the same setup parameters as in \cite{nielQuantumClassicalModeling2018}, which we take as being the ``ground-truth'' benchmark. In these simulations, the electrons can be seen to cool down quicker for higher $\chi_0$, and the distribution functions occupy different fractions of the energy domain throughout the evolution, thus having different requirements on the resolution for the energy grid.

The VQS algorithm maps the dynamics of the quantum state, equation \ref{eq:fokker}, to those of the variational parameters $\theta(t)$ of the ansatz. The mapping is performed using McLachlan’s variational principle (MVP) \cite{mclachlanVariationalSolutionTimedependent1964, endoVariationalQuantumSimulation2020}.
Using the same notation as in \cite{kuboVariationalQuantumSimulations2021},
the unitary quantum circuit produces a wavefunction $|v\{\boldsymbol{\theta}(t)\}\rangle$. However, diffusive processes generally do not preserve the $L_2$ norm, that is, $\partial_t \int |v|^2~\mathrm{d}x \neq 0$. Therefore, it is useful to define an overall variational wavefunction $|\tilde{v}\{\boldsymbol{\theta}(t)\}\rangle = \alpha(t)~|v\{\boldsymbol{\theta}(t)\} \rangle $
where $\alpha(t)$ is a (classical) normalization parameter.
If the temporal resolution is sufficiently high and the ansatz sufficiently expressive, the MVP method evolves the variational parameters such that the correct dynamics are enforced through

\begin{equation}
    \min _{\theta(t)} \bigg\|  \frac{\mathrm{d}}{\mathrm{d} t}|\tilde{v}\{\boldsymbol{\theta}(t)\}\rangle-\hat{L}~|\tilde{v}\{\boldsymbol{\theta}(t)\}\rangle \bigg\| 
\end{equation}

\noindent where $\hat{L}$ is a linear operator that generates the evolution of the system. At each time step, the following matrix equation needs to be solved

\begin{equation}
    M_{k,j} ~\dot{\theta}_j = V_k
\end{equation}

\noindent where

\begin{equation}
    M_{k,j} \equiv \mathrm{Re} \left(  \dfrac{\partial \langle \tilde{v} \{ \theta(t) \} | }{\partial \theta_k} \dfrac{\partial | \tilde{v} \{ \theta(t) \} \rangle }{\partial \theta_j}  \right), ~V_k \equiv  \mathrm{Re} \left( \dfrac{\partial \langle \tilde{v} \{ \theta(t) \} | }{\partial \theta_k} \hat{L} | \tilde{v} \{ \theta(t) \} \rangle \right)
    \label{eq:MVP_Mkj_Vk}
\end{equation}

\noindent are quantities that can be obtained efficiently through dedicated quantum circuits. One advantage of this method, if run on a quantum device, is that one does not need to have access to the full amplitudes of the wavefunction. Instead, only the variational parameters need to be tracked, and their evolution is obtained through efficient measurement of a few observables.
Whereas the $M_{k,j}$ only depend on the chosen ansatz, the $V_k$ elements depend on both the ansatz and the particular Fokker-Planck equation operators (in this case equation \ref{eq:fokker}).
In this work, we classically simulate the retrieval of these quantities through numerical differentiation and integration of the wavefunction using \texttt{Numpy} \cite{harrisArrayProgrammingNumPy2020a}.

\begin{figure}
    \centering
    \includegraphics[width=300pt]{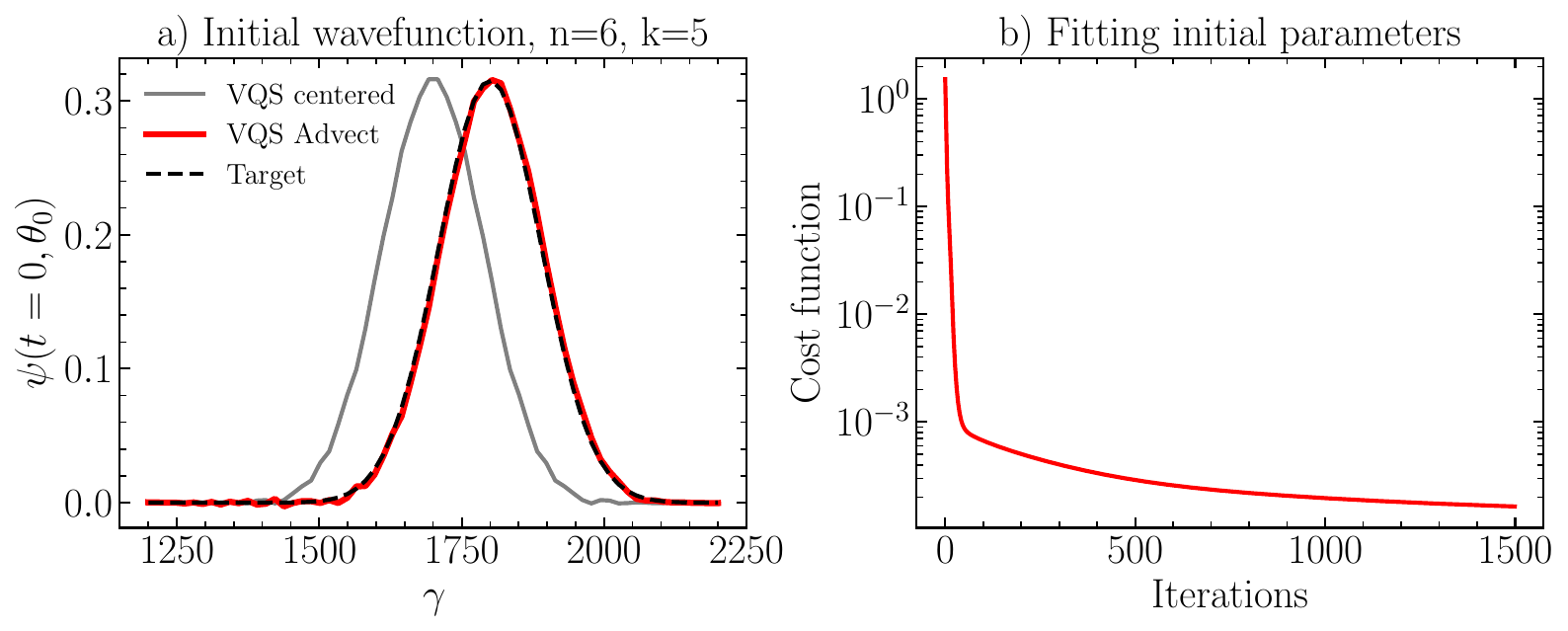}
    \caption{Fitting of the initial wavefunction for $\chi_0=10^{-3}$, $n=6$ qubits and $k=5$ layers of variational parameters. Left: a centered wavefunction is obtained through fitting the variational parameters. Right: evolution of the cost function as the optimizer converges on a good approximation of the target wavefunction.}
    \label{fig:FokkerPlanck_VarQITE_chi3_psi0_cost}
\end{figure}

As previously explained, we first optimize the parameters (except for $\theta_\mu$, which we keep fixed) using a cost function $C = ||\psi(\theta)-\psi_{target}||^2$. We choose a number of iteration steps such that the cost function has decreased significantly and converged such that the quantum variational ansatz produces the centered target distribution (see figure \ref{fig:FokkerPlanck_VarQITE_chi3_psi0_cost}). After this, we apply the advection step by changing the parameter $\theta_\mu$ to match the initial mean value of the distribution on the energy grid.

\begin{figure}
    \centering
    \includegraphics[width=300pt]{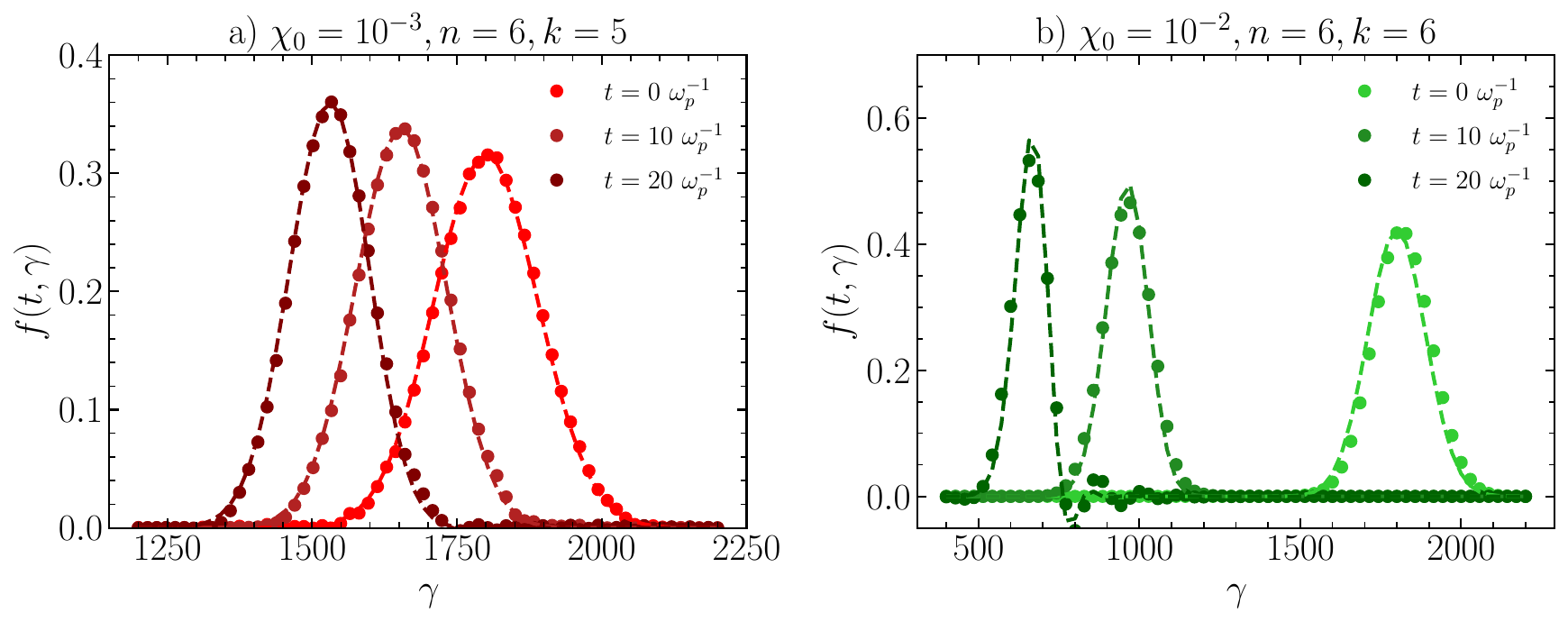}
    \caption{Evolution of electron distribution functions for $\chi_0={10^{-3},10^{-2}}$. Dashed line - PDE solver, Circles - VarQITE. In both cases, $n=6$ qubits, while the number of layers is $k=5,6$ for the two setups. a) simulation results for $\chi = 10^{-3}$. b) simulation results for $\chi = 10^{-2}$ }
    \label{fig:FokkerPlanck_VarQITE_distributions}
\end{figure}

In figure \ref{fig:FokkerPlanck_VarQITE_distributions}, we show the VarQITE simulation of the distribution function for $\chi_0=\{ 10^{-3}, 10^{-2}\}$, and results from solving the PDE with a classical algorithm. We thus show the capability of VarQITE to simulate classical distribution functions relevant to plasma physics.

\begin{figure}
    \centering
    \includegraphics[width=300pt]{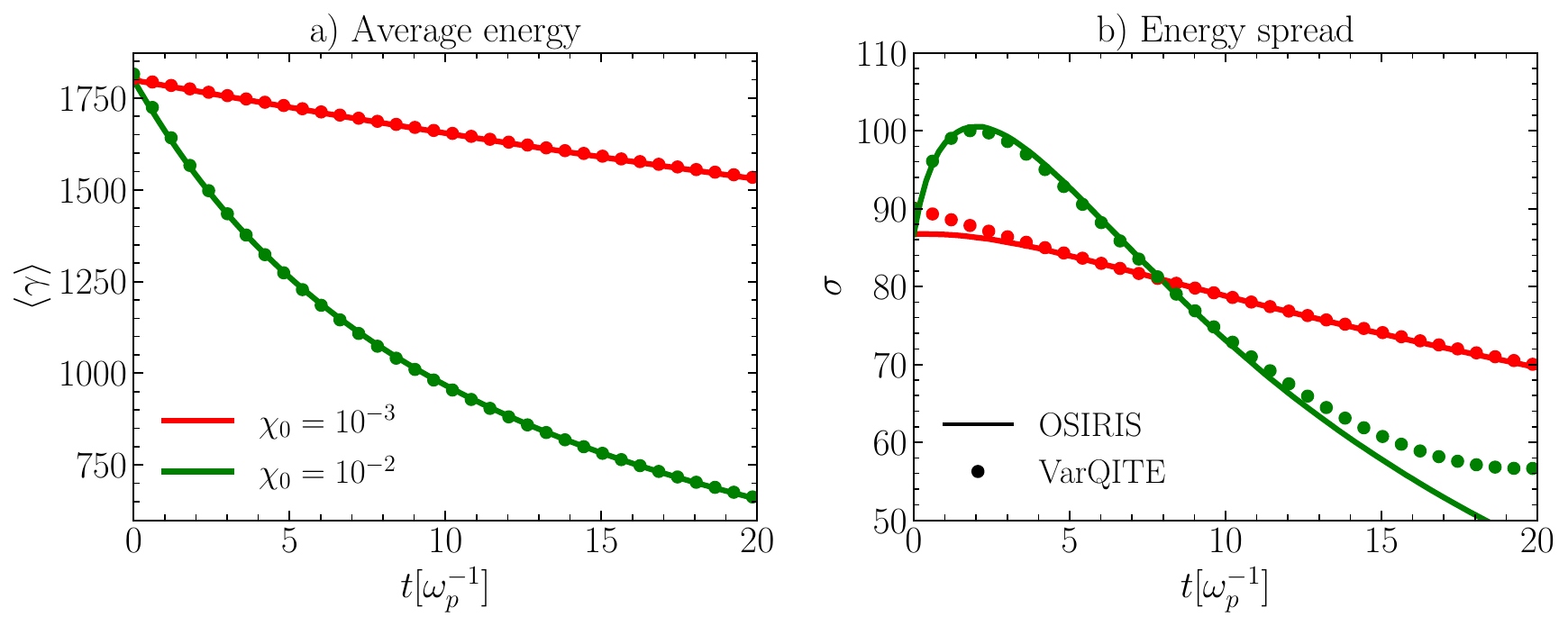}
    \caption{Evolution of the moments of the distribution functions obtained through VarQITE (circles) and compared against \texttt{OSIRIS} (lines). a) the first moment (mean), b) second moment (spread).}
    \label{fig:FokkerPlanck_VarQITE_moments}
\end{figure}

In figure \ref{fig:FokkerPlanck_VarQITE_moments}, we show the evolution of the first two moments of the distribution functions. There is a general agreement of the quantum circuit results with the \texttt{OSIRIS} simulations. There is a small deviation due to finite grid resolution, which can be resolved by increasing the number of qubits.

\begin{figure}
    \centering
    \includegraphics[width=300pt]{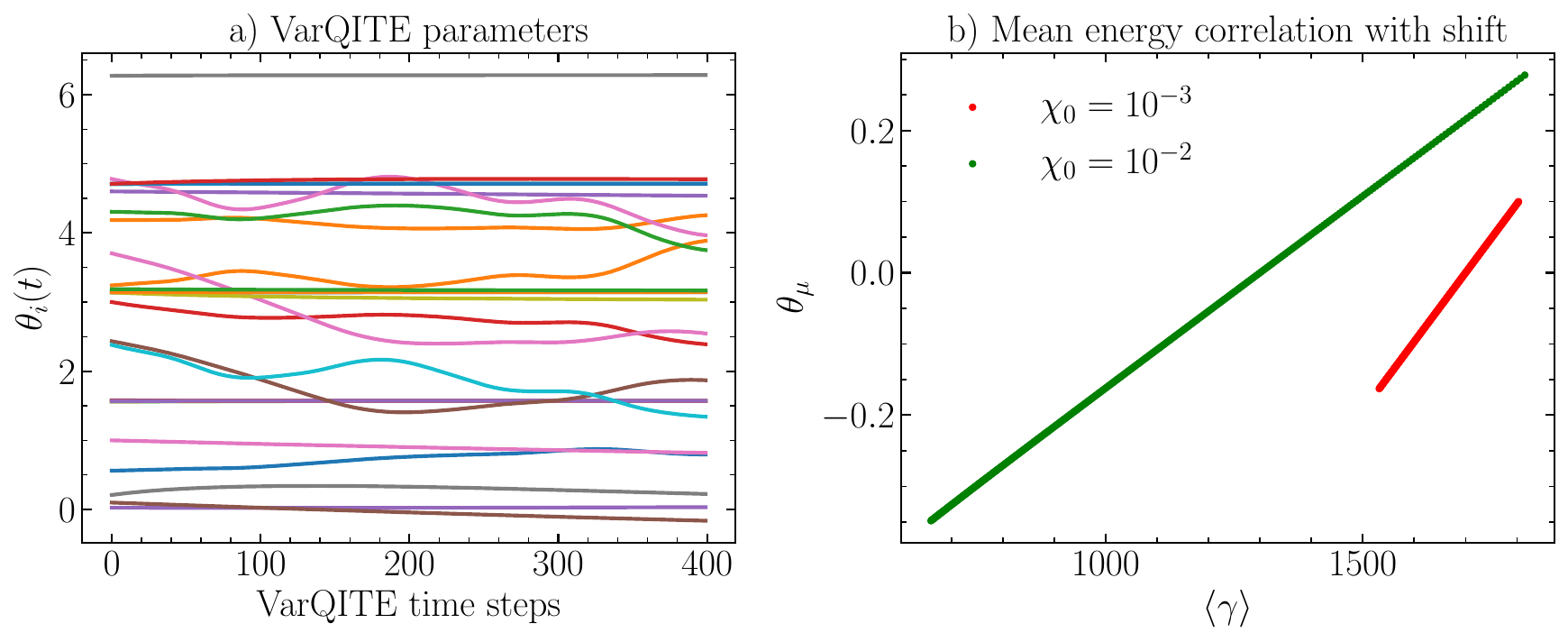}
    \caption{a) The evolution of variational parameters for $\chi_0=10^{-3}$, b) the linear correlation between average energy from the wavefunction and the parameter of the wavefunction translation.}
    \label{fig:FokkerPlanck_VarQITE_parameters_advection}
\end{figure}

In figure \ref{fig:FokkerPlanck_VarQITE_parameters_advection}, we show the evolution of all the $27$ variational parameters for the $\chi_0=10^{-3}$ case. The trajectories are smooth, and some parameters have minimal deviations, which suggests that the ansatz is over-parametrized and can be made more efficient. Figure \ref{fig:FokkerPlanck_VarQITE_parameters_advection} b) shows the values of the mean energy computed from the wavefunction produced by the quantum variational ansatz against the variational parameter responsible for the translation $e^{-i \theta_\mu \partial_x }$, where a linear scaling between the two is visible.

\subsection{A note on the extraction of moments of the distribution function}

The global translation of the wavefunction is correlated with the average energy through a simple rescaling. Therefore, we would not need to measure the mean energy of the wavefunction because we could simply read this value from one of the variational parameters. This is only allowed because of our choice of a translated, even-symmetric wavefunction and would not be viable if the ansatz allowed for skewed distributions.

The average and variance of the energy can be measured efficiently ($n~Z_i$, plus $n$ and $2n$ $Z_iZ_j$ gates, respectively), with a Pauli-string decomposition of the  observables:
\begin{equation}
    \hat{x} = \frac{1}{2}(2^n - 1) \, I - \frac{1}{2} \sum_{i=0}^{n-1} 2^{n - i - 1} Z_i
\end{equation}

\begin{equation}
    \hat{x}^2 = \left( a^2 + \sum_{i=0}^{n-1} b_i^2 \right) \, I + 2a \sum_{i=0}^{n-1} b_i Z_i + 2 \sum_{0 \leq i < j \leq n-1} b_i b_j \, Z_i Z_j    
\end{equation}

with $a = \frac{1}{2}(2^n - 1), b_i = -\frac{1}{2} \times 2^{n - i - 1}$, and $Z_i$ is the Pauli-$Z$ gate applied to qubit $i$. The commutation relation between the individual terms can be used to make the measurement more efficient, and the results of $\langle x\rangle = \langle \psi|\hat{x}|\psi\rangle, \langle x^2\rangle = \langle\psi|\hat{x}^2|\psi\rangle$ can then be linearly rescaled to be compatible with the wavefunction domain in energy $\gamma$.

\section{Conclusions}\label{sc:conclusions}

In this work, we present a quantum-hybrid algorithm approach to Quantum Radiation Reaction, providing similar results when benchmarked against the particle-in-cell (PIC) code \texttt{OSIRIS}.
We have simulated the Fokker-Planck equation describing the stochastic cooling of an electron population in a strong magnetic field using the Variational Quantum Imaginary Time Evolution algorithm (VarQITE), showing good agreement with the results from PIC.
We have developed a new variational ansatz that mimics a Gaussian trial-function, capturing the first and second moments of the distribution functions while requiring fewer parameters than more general variational ansatze.
We have also derived closed-form analytical results for moments of the energy distribution functions, their entropy and autocorrelation functions.

Future steps will include extending this approach to other Fokker-Planck equations of interest (for example, the Kompaneets equation or laser cooling of trapped atoms), to the full Boltzmann equation (nonlocal operator), and to represent the quantum system not as a classical distribution function but as a Fock-state \cite{hidalgoQuantumSimulationsStrongfield2024}, where the dynamics would be naturally unitary.
Additionally, using specialized circuits for loading specific distribution functions as part of the variational ansatz (Gaussian, Uniform, Chi-squared) could lead to a reduced number and more interpretable parameters, which would speedup the VarQITE algorithm and possibly mitigate the barren-plateau problem.

Our work can contribute to further development of both classical diffusive plasma physics and 
to simulate the fully quantum nature of plasma interactions, namely the transition from semi-classical to nonperturbative SFQED regimes.

\section{Data and Code availability statement}\label{sc:codeavailability}

The data that support the findings of this study are openly
available at the following URL/DOI: \url{https://github.com/OsAmaro/QuantumFokkerPlanck}. The repository includes notebooks explaining the classical PDE solver, the VarQITE approach, the analytical models and input decks for \texttt{OSIRIS} simulations.

\section{Acknowledgments}\label{sc:acknowledgments}


The authors thank the following people for fruitful discussions / proofreading parts of the manuscript:
Mr. José Mariano on Monte-Carlo sampling,
Mr. Lucas Ansia on classical simulation of the Fokker-Planck equation, 
Mr. Bernardo Barbosa on Quantum Radiation Reaction in SFQED,
Mr. Pablo Bilbao on plasma dynamics in strong magnetic fields,
Mr. Gabriel Almeida on expressibility and complexity of variational quantum circuits,  
Mr. Anthony Gandon on efficient Pauli-String decomposition of Hamiltonians for variational quantum circuits,
Mr. Efstratios Koukoutsis on Lindbladian/dissipative quantum simulations, and
Mr. Diogo Cruz on quantum algorithms for PDE solvers.

This work was supported by the Portuguese Science Foundation (FCT) Grants No. PTDC/FIS-PLA/3800/2021, UI/BD/153735/2022.
This work has been carried out within the framework of the EUROfusion Consortium, funded by the European Union via the Euratom Research and Training Programme (Grant Agreement No 101052200 — EUROfusion). Views and opinions expressed are however those of the authors only and do not necessarily reflect those of the European Union or the European Commission. Neither the European Union nor the European Commission can be held responsible for them.




Simulations were performed at the IST cluster (Lisbon, Portugal).
We have used the quantum frameworks Qiskit \cite{qiskit2024} and PennyLane \cite{bergholmPennyLaneAutomaticDifferentiation2022}.

\appendix

\section{Applying the Variational Principle to the Heat equation}\label{appendix:heat}

In this appendix, we provide a step-by-step application of McLachlan's Variational Principle (MVP) to the simplest diffusive PDE, the heat equation. 
Similar to the Classical Variational Principle used in Quantum Mechanics to obtain approximate Ground-States of a system, the MVP for imaginary or real-time evolution is useful both numerically and analytically.

The 1D homogeneous heat equation

\begin{equation}
    \partial_t u = k ~\partial_{xx} u
\end{equation}

\noindent with thermal diffusivity coefficient $k$, admits a kernel solution (when $u(t=0,x)=\delta(x)$) 

\begin{equation}
    u(t,x) = \dfrac{1}{\sqrt{4 \pi k t}} \exp \left( -\dfrac{x^2}{4 k t} \right)
    \label{eq:heat_solution}
\end{equation}

\noindent from which analytical solutions to more general initial conditions can be constructed.

This suggests using an ansatz 

\begin{equation}
    v(A,\sigma) = A \exp \left( -\dfrac{x^2}{2\sigma^2} \right), ~\partial_t v = \hat{L}_H~v
    \label{eq:heat_ansatz}
\end{equation}

\noindent with  $\hat{L}_H$ the linear operator generating the evolution of the system (in this case, it is the Laplacian $\partial_{xx}$), and variational parameters $\theta_i(t)=(A(t),\sigma(t))$, for the analytical solution of the MVP. We take the thermal diffusivity to be $k=1$ for simplicity. Since this function is $L_1$-normalized (that is, $\int v ~\mathrm{d}x = 1$), the normalization $A$ will be correlated and a function of the parameter $\sigma$.

The following derivatives and integrals are used in the calculations

$$
    \partial_A v = A^{-1}~v, ~\partial_\sigma v = x^2~\sigma^{-3} ~v, ~\partial_{xx} v = v~(x^2-\sigma^2)~\sigma^{-4}
$$

\begin{equation}
    \int v ~\mathrm{d}x = A \sqrt{2\pi} \sigma, \int v^2 ~\mathrm{d}x = A^2 \sqrt{\pi} \sigma
\end{equation}

$$
    \int x^2 v^2 ~\mathrm{d}x = A^2 \sqrt{\pi} \sigma^3 /2, \int x^4 v^2 ~\mathrm{d}x = 3 A^2 \sqrt{\pi} \sigma^5 /4
$$

From the MVP equations \ref{eq:MVP_Mkj_Vk}, we need to compute the matrix elements $\int (\partial_k v )(\partial_j v )~\mathrm{d}x$, which require either the $\int v^2~\mathrm{d}x$ or $\int x^2~v^2~\mathrm{d}x$ integrals. We obtain

$$
M_{A,A} = \sqrt{\pi} \sigma, ~M_{A,\sigma} = M_{\sigma,A} = \sqrt{\pi} A/2, ~M_{\sigma,\sigma} = 3\sqrt{\pi} A^2 /(4\sigma), 
$$

\noindent For the column vector, we need to compute $\int (\partial_k v) (\hat{L}_H v)~\mathrm{d}x$. We obtain

$$
V_A = -A \sqrt{\pi} / (2\sigma), ~V_\sigma = A^2 \sqrt{\pi} / (4\sigma^2)
$$

\begin{figure}
    \centering
    \includegraphics[width=300pt]{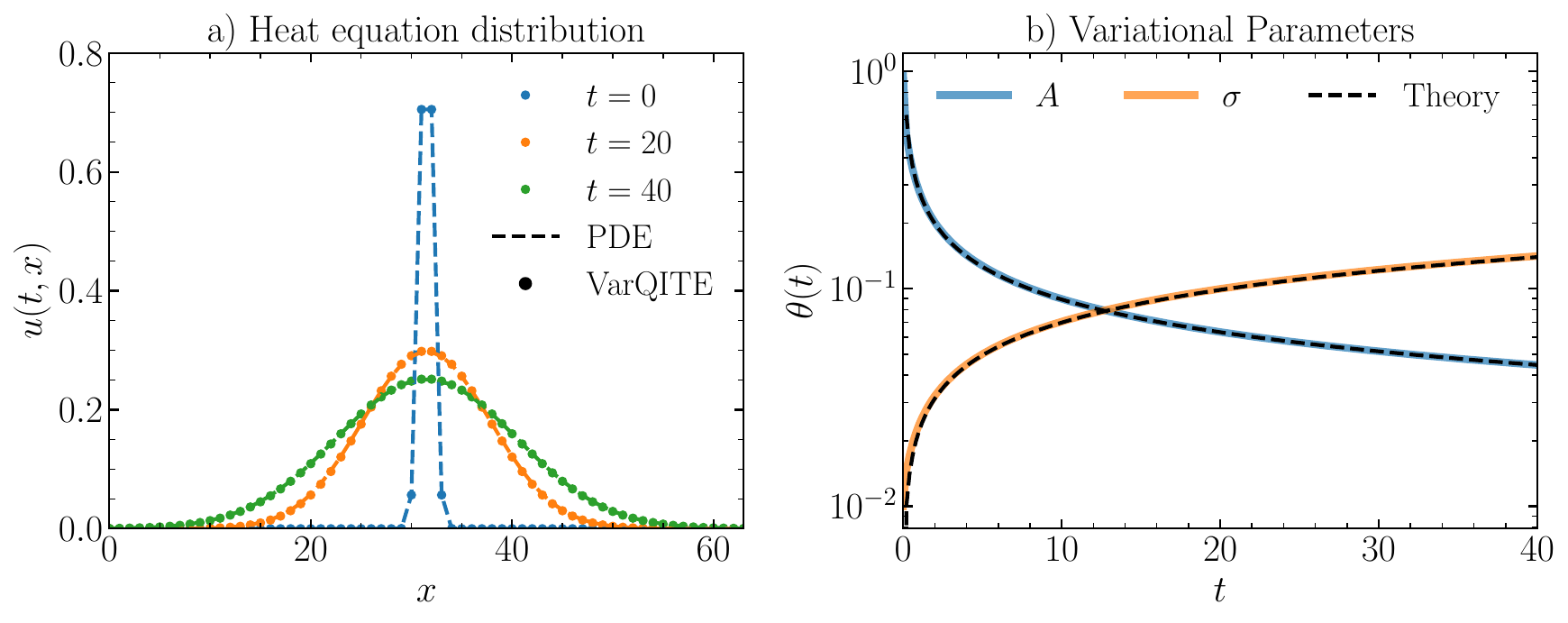}
    \caption{Numerical solution of the 1D Heat equation. a) Snapshots of the distribution $u(t,x)$ from a PDE solver and numerical evolution of a Gaussian ansatz in the MVP. b) Evolution of the variational parameters: color - numerical MVP, dashed - analytical solution to the MVP equations \ref{eq:heat_MVP_solution}.}
    \label{fig:Heat_analytical_MVP}
\end{figure}

\noindent We now need to solve the matrix equation $M_{k,j} \dot{\theta}_j = V_k$. Factors of $\sqrt{\pi}$ will cancel out.

\begin{equation}
    \begin{bmatrix}
        \sigma & A/2 \\
        A/2 & 3A^2/(4\sigma)
    \end{bmatrix}
    \begin{bmatrix}
        \dot{A} \\ \dot{\sigma}
    \end{bmatrix} = 
    V_k \leftrightarrow
    \begin{bmatrix}
        \dot{A} \\ \dot{\sigma}
    \end{bmatrix} = \begin{bmatrix}
        3/(2\sigma) & -1/A \\
        -1/A & 2\sigma/A^2
    \end{bmatrix}
    \begin{bmatrix}
        -A/(2\sigma) \\ A^2 /(4\sigma^2)
    \end{bmatrix} = \begin{bmatrix}
        -A/\sigma^2 \\ 1/\sigma
    \end{bmatrix}
\end{equation}

\noindent where we explicitly computed the matrix inverse of $M_{k,j}$. The second equation (for $\sigma$) is separable. We can then replace $\sigma(t)$ in the first equation to obtain the amplitude $A(t)$. The parameters thus become

\begin{equation}
    \sigma(t) = \sqrt{2t + \sigma_0^2}, ~A(t) = A_0 \sigma_0/\sqrt{2 t + \sigma_0^2}
    \label{eq:heat_MVP_solution}
\end{equation}

\noindent where $\sigma_0, A_0$ are the initial spread and amplitude, respectively. This is precisely the exact analytical solution \ref{eq:heat_solution}. In figure \ref{fig:Heat_analytical_MVP}, we show a comparison between a standard PDE solver solution of the heat equation, a numerical solution of the MVP equations using the ansatz \ref{eq:heat_ansatz} (without a quantum circuit), and the analytical solution \ref{eq:heat_MVP_solution}.
The numerical grid had $2^6$ cells, the number of timesteps was $200$ for a maximum simulation time of $40$, and a small initial spread of the distribution function $\sigma_0=0.01$.

\section{The Fokker-Planck equation}\label{appendix:fokkerplanck}

\subsection{Review of theoretical study of Radiation Reaction}

In this section we review past works on radiation reaction, introduce the main parameters that describe the interaction of electrons with an intense magnetic field, and derive analytical solutions for some observables of interest.

In the past decades, the theoretical modelling of electron radiation losses in intense fields has been extensively studied.
In \cite{shenEnergyStragglingRadiation1972}, authors derive the transport partial differential equation (PDE) for electrons in a constant uniform magnetic field.
Analytical results of multi-photon scattering of electrons in periodic structures (e.g. oriented-crystals) have been derived in \cite{khokonovCascadeProcessesEnergy2004, bondarencoMultiphotonEffectsCoherent2014}, however without simple explicit closed-form expressions for the observables, nor immediate application in electron-laser scattering.
In \cite{lobetModelingRadiativeQuantum2016}, the coupled dynamics of photons and leptons, including radiation reaction and pair production in a constant magnetic field, are studied numerically.
In \cite{vranicQuantumRadiationReaction2016, ridgersSignaturesQuantumEffects2017, nielQuantumClassicalModeling2018} the authors derive the Ordinary Differential Equations (ODEs) for the first two momenta of the energy distributions of electrons interacting with a laser pulse, but don't present their corresponding closed-form explicit solutions.
In \cite{nielQuantumClassicalModeling2018}, the authors show connections between the different regimes of radiation reaction and the corresponding particle trajectories and stochastic PDEs.
In \cite{artemenkoGlobalConstantField2019}, the electron transport PDE is solved numerically for different magnetic field values using the ``Global Constant Field Approximation'' to map between a constant magnetic field setup and a laser pulse with a finite duration envelope.
In \cite{jirkaReachingHighLaser2021}, the authors derive the average energy decay in a pulsed Plane Wave laser in the highly ``nonlinear quantum'' high $\chi$ regime (the quantum nonlinear parameter, $\chi$, is defined in section \ref{sc:analytical}).

More recently, some papers have contributed with new analytical results for the study of classical (CRR) and quantum (QRR) radiation reaction regimes.
In \cite{bilbaoRadiationReactionCooling2023}, authors derive analytically the evolution of an electron momentum ``ring distribution'' in a maser setup, relevant for astrophysical plasma scenarios. The formulas are derived in the CRR regime using the ``method of lines'' technique but appear to hold even for $\chi \sim 0.1$, and suggest that this class of ring distributions is a ``global attractor''.
In \cite{zhangQuantumSplittingElectron2023}, the authors derive approximate analytical formulas for the evolution of a lepton distribution function in a constant magnetic field during multiphoton scattering in the QRR $\chi < 1$ regime. The authors identify and analytically characterise the ``quantum peak splitting'', which occurs when an initially peaked distribution function evolves into a doubly peaked distribution.
In \cite{bulanovEnergySpectrumEvolution2024}, the authors derive CRR closed-form expressions for distribution functions for a set of initial conditions of interest, again using the ``method of lines'', and find an equilibrium (time-independent) solution for the QRR Fokker-Planck equation of an electron beam in a LWFA setup.
In \cite{kostyukovShorttermEvolutionElectron2023a}, the authors derive short-time evolution expressions for the electron wave packet in a strong EM field, including the expectation value of the electron spin. The approach relies on Volkov functions and the Dyson-Schwinger equation and naturally leads to the damping (radiation reaction) of the wave packet.
In \cite{torgrimssonQuantumRadiationReaction2024,torgrimssonQuantumRadiationReaction2024a} the author derives explicit analytical time-dependent electron distribution functions in momentum and spin through resummation techniques. Following a similar approach,
in \cite{blackburnAnalyticalSolutionsQuantum2024}, the author derives explicit analytical QRR formulas for the final electron mean energy and spread as functions of initial electron energy, laser $a_0$ and pulse duration. These expressions are derived in the low $\chi$ regime but remain approximately valid for a larger range of parameters.

\subsection{Regimes of Radiation Reaction}

Here we follow the notation of \cite{vranicQuantumRadiationReaction2016}.
When $\chi \ll 1$, the equation for energy loss can be derived from the Landau-Lifshitz equation, leading to
\begin{equation}
    \dfrac{\mathrm{d}\gamma}{\mathrm{d}t} = -c_{rr} ~\gamma^2
\end{equation}

\noindent with $c_{rr} \equiv 2 ~\omega_c e^2 b_0^2/(3 m c^3)$ a constant defining the effect of Classical Radiation Reaction. In the setup considered, 3 physical parameters completely describe the dynamics: the initial electron energy $\gamma_0$, the external magnetic field $B$, and the evolution time $t$. The solution for the energy is 

\begin{equation}
    \gamma_c(t)\sim \gamma_0/(1+c_{rr}~t~\gamma_0)
    \label{eq:gtBconstSol}
\end{equation}

\noindent This scaling for the average energy can be corrected in the $\chi \lesssim 1$ regime.

From a distribution point of view, the relativistic Vlasov equation for the electron distribution function $f$ is modified. Radiation Reaction no longer preserves phase-space volume, corresponding to the operator on the RHS of the equation

\begin{equation}
    \frac{\mathrm{d} f(t, \vec{x}, \vec{p})}{\mathrm{d} t} = \frac{\partial f}{\partial t} + v \cdot \nabla_x ~f + F \cdot \nabla_p f
    = \mathcal{C}[f_e]
    \label{eq:vlasov}
\end{equation}

\noindent where $\mathcal{C}$ is a collision operator whose form depends on the model/regime of radiation reaction considering \cite{nielQuantumClassicalModeling2018}. In our simplified setup, the coordinate gradient and the force terms vanish, so the only term of interest is the collision operator.

For higher values of $\chi$, the emission of photons becomes stochastic, in general leading to a spread in energy. Conceptually, the electrons can be thought of as being slightly scattered through a cloud of photons of the field, where each photon has an energy much lower than the electron's.

The probability of emission of a photon with $\chi_\gamma$ by an electron with $\chi$ per unit time, also known as the  nonlinear Compton Scattering (nCS) differential cross section/rate, is 
\begin{multline}
    \frac{\mathrm{d}^{2} N_\gamma}{\mathrm{d} t \ \mathrm{d} \chi_\gamma} (\chi,\chi_\gamma) =\frac{\alpha m c^{2}}{\sqrt{3} \pi \hbar \gamma \chi}\left[ \left(1-\xi +\dfrac{1}{1-\xi} \right) K_{2 / 3}(\tilde{\chi})  - \int_{\tilde{\chi}}^{\infty}  \ K_{1 / 3}(x) ~\mathrm{d} x \right] 
    \label{eq:nCS_diff}
\end{multline}
\noindent with $\xi  \equiv  \chi/\chi_\gamma$, $\tilde{\chi} \equiv 2\xi/(3\chi(1-\xi))$, and $K_n$ is the modified Bessel function of second kind \cite{abramowitz1964handbook}.
This rate in $\chi_\gamma$ can be mapped to a rate in particle energy $\mathrm{d}^2 N_\gamma/\mathrm{d}t \mathrm{d}\gamma_\gamma~(\gamma,\gamma_\gamma)$.

The Fokker-Planck equation (FP) can be seen as an extension of the standard Vlasov equation to kinetically and stochastically model collisions between plasma species and laser-plasma interaction. An important application is simulating the energy loss of an electron beam as it interacts with electromagnetic fields \cite{neitzStochasticityEffectsQuantum2013,vranicQuantumRadiationReaction2016}. 
In this case, the particle distribution evolves through

\begin{equation}
    \frac{\partial f(t, \vec{p})}{\partial t}=\frac{\partial}{\partial p_{l}}\left[-\mathcal{A}_{l} f+\frac{1}{2} \frac{\partial}{\partial p_{k}}\left(\mathcal{B}_{l k} f\right)\right]
    \label{eq:fokker_momentum}
\end{equation}

with $\mathcal{A}_{l} \equiv \int q_{l} w(\vec{p}, \vec{q}) \mathrm{d}^{3} \vec{q}$, $\mathcal{B}_{l k} \equiv \int q_{l} q_{k} w(\vec{p}, \vec{q}) \mathrm{d}^{3} \vec{q}$, the drift and diffusion coefficients respectively, and $ w(\vec{p}, \vec{q})~\mathrm{d}^3\vec{p}$ is the probability per unit time of momentum change of the electron  $\vec{p}\rightarrow\vec{p}-\vec{q}$, with $\vec{q}$ the momentum of the photon, and where Einstein index contraction was assumed.

In the co-linear approximation of radiation emission, $\chi_\gamma/\chi \sim \hbar k/(\gamma m c)$, this problem becomes essentially 1D, with drift and diffusion coefficients

\begin{equation}
    \mathcal{A} \equiv \dfrac{\gamma m c}{\chi} \int_0^{\chi} \chi_\gamma ~\dfrac{\mathrm{d}^2 N_\gamma}{\mathrm{d}t\mathrm{d}\chi_\gamma} ~ \mathrm{d}\chi_\gamma, ~\mathcal{B} \equiv \dfrac{(\gamma m c)^2}{\chi^2} \int_0^{\chi} \chi_\gamma^2 ~\dfrac{\mathrm{d}^2 N_\gamma}{\mathrm{d}t\mathrm{d}\chi_\gamma} ~ \mathrm{d}\chi_\gamma
    \label{eq:FP_AB_collinear}
\end{equation}

\noindent In the regime of $\chi \ll 1$, the $\mathcal{A}$ and $\mathcal{B}$ coefficients have polynomial approximations \cite{vranicQuantumRadiationReaction2016}

\begin{equation}
    \begin{split}
        \mathcal{A} \sim \dfrac{2}{3} \dfrac{\alpha m^2 c^3}{\hbar} \chi^2 \propto \gamma^2, ~\mathcal{B} \sim \dfrac{55}{24 \sqrt 3} \dfrac{\alpha m^3 c^4}{\hbar} \gamma ~ \chi^3 \propto \gamma^4
    \end{split}
    \label{eq:FP_AB_app2}
\end{equation}

%
The Fokker-Planck equation is sparse (only 2 local, differential operators), while the Boltzmann equation is nonlocal, and its corresponding Hamiltonian evolution matrix is dense. Consequently, both numerical (either a PDE solver or Monte-Carlo sampling) and analytical solutions of the latter equation are often more challenging than for the former.

\subsection{Entropy and Autocorrelation}

Having derived the evolution of the spread \ref{eq:Blackburn_spread}, and assuming a Gaussian functional form of the distribution function, the Shannon entropy of the electron beam evolves as $S(t) = -\int f \log(f) ~\mathrm{d}\gamma \sim \log(\sigma(t)) + c^{te}$ up to an additive constant (which we choose such that the simulation and theory curves match at late times).
Since this is a monotone function of the spread, it has the same qualitative behavior and peaks at $\log(\sigma_{\max})$.
From a physics point-of-view, the entropy changes due to a competition between the drift and diffusion processes. At late times, the entropy of the electrons decreases, which can be interpreted as a transfer of entropy to the radiation that is being emitted.

One can also compute an approximate auto-correlation function from the classical solution $\gamma_c(t) \sim \gamma_0/(1+2R_c t/3)$ as

\begin{equation}
    \begin{split}
    g(\tau) \equiv \dfrac{\langle \gamma(t) \gamma(t+\tau) \rangle}{\langle \gamma^2(t) \rangle} = \int_0^\infty \dfrac{1}{(1+2 R_c t/3)(1+2 R_c (t+\tau)/3)} \mathrm{d}t \\
    \propto \dfrac{\log(1+2 R_c \tau/3)}{2 R_c \tau/3}
    \end{split}
    \label{eq:autocorr1st}
\end{equation}

\noindent where normalization $g(0) = 1$ is enforced. This quantifies how correlated two electron trajectories are on average if measured with a time delay of $\tau$. 
Using instead the 2nd order expansion for the average energy (that is, the full equation \ref{eq:Blackburn_average}) leads to a more accurate closed-form solution, albeit with more terms (not shown here). Considering the stochastic component of the trajectories would require computing expectation values of nonlinear functions of Wiener processes $W(t), W(t+\tau)$.
The solution to the Fokker-Planck equation $f(t,\gamma)$ does not give us information on the auto-correlation directly, and a marginal distribution $f(t_1,\gamma_1; t_2, \gamma_2)$ would have to be obtained.

\begin{figure}
    \centering
    \includegraphics[width=300pt]{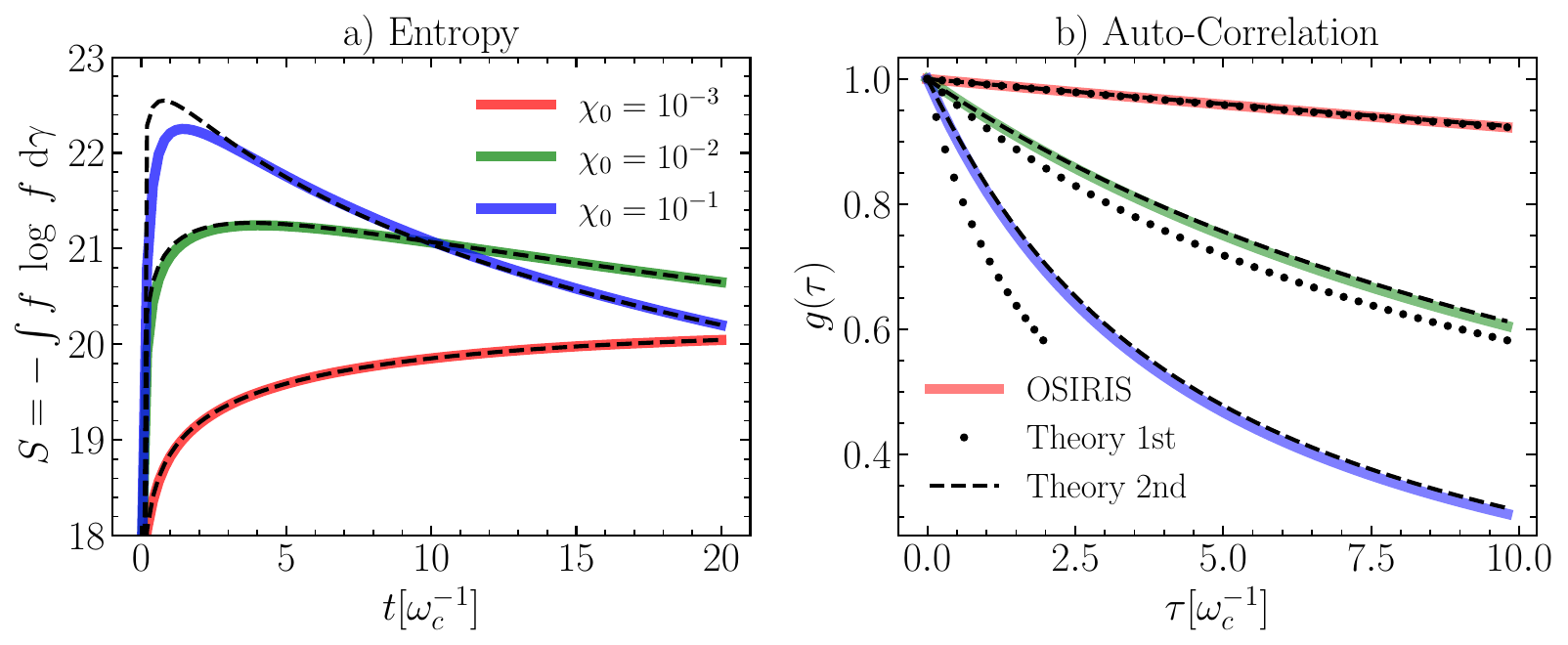}
    \caption{a) The evolution of the distribution function entropy, b) auto-correlation function, where dotted lines represent equation \ref{eq:autocorr1st} and
    dashed lines are the result of using \ref{eq:Blackburn_average} in $g(\tau)$. In both figures, lines in color represent results from \texttt{OSIRIS} simulations.}
    \label{fig:FokkerPlanck_entropy_autocorrelation}
\end{figure}

In figure \ref{fig:FokkerPlanck_entropy_autocorrelation}, we show the comparison between these two quantities $S(t)$, $g(\tau)$ from \texttt{OSIRIS} simulations computed from distribution histograms and individual particle trajectories, and the approximate analytical results. Again, for $\chi_0=10^{-1}$ (blue curve), the $t$-axis is again multiplied by a factor of $2$ to have the same range as the other curves.

\subsection{McLachlan’s Variational Principle approach}

Here, we apply the technique described in \ref{appendix:heat} to the Fokker-Planck equation.
Since the number of particles ($L_1$ norm) is conserved and we are mostly interested in the first moments of the distribution function $\mu=\int \gamma f~\mathrm{d}\gamma$, $\sigma^2 = \int (\gamma-\mu)^2 f~\mathrm{d}\gamma$, we enforce this into the ansatz

\begin{equation}
    v(\mu,\sigma) = \dfrac{1}{\sqrt{2\pi \sigma^2}} \exp \left( -\dfrac{(\gamma-\mu)^2}{2\sigma^2} \right), ~\partial_t v = \hat{L}_{FP}~v = -\partial_\gamma(a ~\gamma^2 ~v) + 0.5 ~\partial_{\gamma \gamma} (b~ \gamma^4~ v)
    \label{eq:fokkerplanck_MVP_ansatz}
\end{equation}

\noindent with $\hat{L}_{FP}$ the linear operator generating the dynamics, and variational parameters $\theta_i(t)=(\mu(t),\sigma(t))$, for the analytical solution of the MVP.
We could have used an ansatz with the amplitude $A(t)$, which would have led to a 3x3 matrix system of equations.

The following derivatives will be used in the following calculations

\begin{equation}
    \dfrac{\partial_\mu v}{v} = \dfrac{\gamma-\mu}{\sigma^2}, ~\dfrac{\partial_\sigma v}{v} = \dfrac{(\gamma-\mu)^2-\sigma^2}{\sigma^3}
\end{equation}

From the MVP equations \ref{eq:MVP_Mkj_Vk}, we need to compute the matrix elements $\int (\partial_k v )(\partial_j v ) \mathrm{d}\gamma$, which require either the $\int (\gamma-\mu)^2 ~v^2 ~\mathrm{d}\gamma$ or the $\int (\gamma-\mu)^4~v^2 ~\mathrm{d}\gamma$ Gaussian integrals. In general, the change of variables $(\gamma-\mu)/\sigma \rightarrow \gamma$ simplifies the calculations. We obtain

$$
M_{\mu \mu} = 1/(4 \sqrt{\pi} \sigma^3), ~M_{\sigma \sigma} = 3 / (8 \sqrt{\pi} \sigma^3 ), ~M_{\mu \sigma} = M_{\sigma \mu} = 0
$$

\noindent where the cross-term is null due to the odd-symmetry of the integrand function. The diagonal structure of this matrix simplifies the calculations considerably. The inverse matrix becomes 

\begin{equation}
    M_{kj}^{-1} = 4 \sqrt{\pi} \sigma^3 \begin{bmatrix}
        1 & 0 \\
        0 & 2/3
    \end{bmatrix}
\end{equation}

For the column vector, we obtain

$$
V_\mu = \dfrac{a(-2\mu^2+\sigma^2)}{8 \sqrt{\pi} \sigma^3} -\dfrac{b}{\sigma^3} (0.423142 \mu^3 + 0.211571 \mu \sigma^2)
$$

$$
V_\sigma = -\dfrac{3 a \mu}{4 \sqrt{\pi} \sigma^2} + b \left( -0.343803+0.105786 \dfrac{\mu^4}{\sigma^4} - 0.95207 \dfrac{\mu^2}{\sigma^2} \right)
$$

Applying the inverse matrix $M_{kj}^{-1}$ to the column vector $V_k$, we obtain

\begin{equation}
    \begin{bmatrix}
        \dot{\mu} \\ \dot{\sigma}
    \end{bmatrix} = \begin{bmatrix}
        \underline{-a \mu^2} + 3 b \mu^3 + 0.5 a \sigma^2 + 1.5 b \mu \sigma^2 \\
        ~\underline{-2 a \mu \sigma + b \mu^4 /(2 \sigma)} - 4.5 b \mu^2 \sigma - 1.625 \sigma^3 b
    \end{bmatrix}
\end{equation}

Changing notation $2 \alpha_{rr} = a$ and $b \mu^4 = B/(m^2 c^2)$,
the underlined terms, which are leading order assuming $\mu \gg \sigma$, recover the results of equations 8 and 14 from the reference \cite{vranicQuantumRadiationReaction2016}. 

\begin{figure}
    \centering
    \includegraphics[width=300pt]{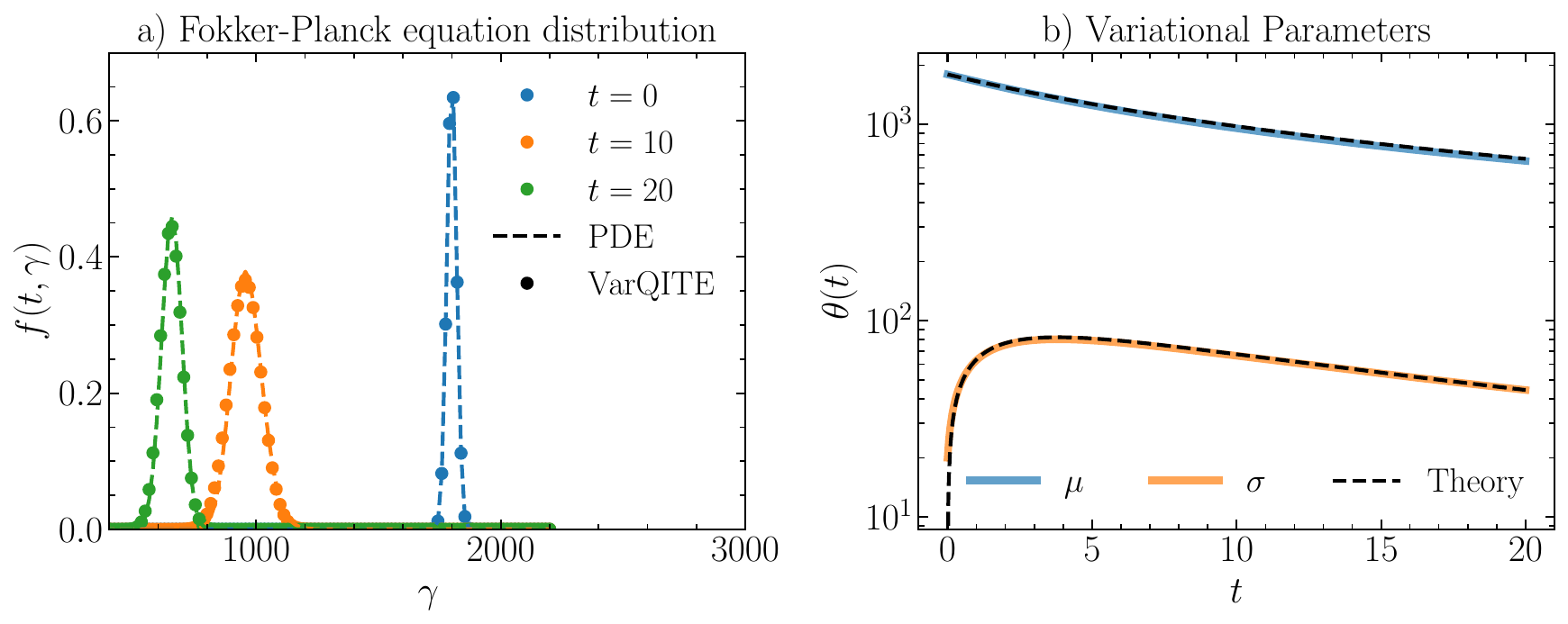}
    \caption{Numerical simulation of the Fokker-Planck equation with $\chi_0=10^{-2}$. a) The evolution of the distribution function $f(t,\gamma)$. The dashed line represents the result from a standard PDE solver, while circles represent the numerical integration of the MVP equations. b) The evolution of the moments. The colored lines represent the moments of the distribution function obtained from the PDE solver, while dashed lines represent the analytical results of equations \ref{eq:Blackburn_average} and \ref{eq:Blackburn_spread}.}
    \label{fig:FokkerPlanck_analytical_MVP}
\end{figure}

In figure \ref{fig:FokkerPlanck_analytical_MVP}, we show a comparison between a PDE solver solution of the Fokker-Planck equation, a numerical solution of the MVP equations using the ansatz \ref{eq:fokkerplanck_MVP_ansatz} (without a quantum circuit), and the analytical solution \ref{eq:Blackburn_average}.
The numerical grid had $2^7$ cells, the number of timesteps was $900$ for a maximum simulation time of $20$, $\chi_0=10^{-2}$, and initial momenta $\mu_0=1800, \sigma_0=20$.

We have thus shown the potential of MVP for retrieving the distribution moments in Quantum Radiation Reaction.

\bibliographystyle{jpp}
\bibliography{bibtex, bibtex_facilities}

\end{document}